\documentclass[prd,twocolumn,showpacs]{revtex4}
\usepackage{hyperref}
\usepackage{epsfig}
\usepackage{url}
\usepackage{amsmath}
\usepackage{amssymb}
\usepackage[pdftex]{color}

\definecolor{Black}{rgb}{.0,.0,.0}
\definecolor{Red}{rgb}{1.,0.,0.}

\begin{document}

\title{Large scale structure simulations of inhomogeneous LTB void models}

\author{David Alonso$^1$, \ Juan Garc\'{\i}a-Bellido$^{1,2}$, \ Troels Haugb{\o}lle$^{3,4,1}$ \ and \ Juli\'{a}n Vicente$^1$} 
\affiliation{$^1$\,Instituto de F\'{\i}sica Te\'{o}rica UAM-CSIC,
Universidad Aut\'{o}noma de Madrid, Cantoblanco, 28049 Madrid, Spain,\\
$^2$\,D\'ep.~Physique Th\'eorique, Univ.~Gen\`eve, 24 quai Ernest 
Ansermet, CH--1211 Gen\`eve 4, Switzerland\\
$^3$\,Niels Bohr International Academy, Niels Bohr Institute, Blegedamsvej 17, 2100 Copenhagen, Denmark\\
$^4$\,Department of Physics and Astronomy, University of Aarhus, DK-8000, Aarhus, Denmark}

\begin{abstract}
We perform numerical simulations of large scale structure evolution in
an inhomogeneous Lema\^itre-Tolman-Bondi (LTB) model of the
Universe. We follow the gravitational collapse of a large underdense
region (a void) in an otherwise flat matter-dominated Einstein-deSitter
model. We observe how the (background) density
contrast at the centre of the void grows to be of order one, and show
that the density and velocity profiles follow the exact non-linear LTB
solution to the full Einstein equations for all but the most extreme voids.
This result seems to contradict previous claims that fully relativistic codes
are needed to properly handle the non-linear evolution of large scale structures, 
and that local Newtonian dynamics with an explicit expansion term is not adequate. 
We also find that the (local) matter density contrast grows with the scale factor 
in a way analogous to that of an open universe with a value of the matter density
$\Omega_M(r)$ corresponding to the appropriate location within the void.
\end{abstract}
\pacs{98.80.Cq\hspace{\stretch{1}}IFT-UAM/CSIC-10-54}

\date{15 October, 2010}

\maketitle

\section{Introduction}

Distant supernovae appear dimmer than expected in a purely matter-dominated
homogeneous and isotropic FRW universe. The currently favoured
explanation of this dimming is the late time acceleration of the universe 
due to an energy component that acts like a repulsive force. 
The nature of the so-called Dark Energy responsible for the apparent 
acceleration is completely unknown.  Observations seem to suggest that 
it is similar to Einstein's cosmological constant, but there is inconclusive 
evidence~\cite{Komatsu:2010fb}. 
In the meantime, our realization that the universe around us is 
far from homogeneous, since there are large superclusters and huge voids 
across our largest galaxy catalogs~\cite{Gott:2003pf}
has triggered the study of alternatives
to this mysterious energy. Since the end of the nineties it has been suggested
by various groups~\cite{LTBvoid,GBH:2008a} that an isotropic but inhomogeneous 
Lema\^itre-Tolman-Bondi universe could also induce an apparent dimming 
of the light of distant supernovae, in this case due to local spatial gradients 
in the expansion rate and matter density, rather than due to late time acceleration. 
There is nothing wrong or inconsistent, apart from philosophical prejudices, 
with the possibility that we live close to the centre of a gigaparsec-sized void. 
Such a supervoid may indeed have been observed as the CMB cold spot~\cite{Cruz} 
and somewhat smaller voids
have been seen in the local galaxy distribution~\cite{Frith:2003,Granett:2008}.
If a local void had the size and depth of a void responsible for the cold spot, 
i.e.~$r_0 \sim 2$ Gpc and 
$\Omega_M \sim 0.2$ within a flat Einstein-de Sitter universe, it would be 
consistent with local observations~\cite{Tully:2007tp,Kashlinsky:2008a}, 
and could account for the supernovae
dimming, together with the observed baryon acoustic oscillations and
CMB acoustic peaks, the age of the universe, local rate of expansion,
etc.~\cite{GBH:2008a,GBH:2008b,GBH:2008c,Biswas:2010xm,Clarkson:2010ej,Moss:2010jx,Yoo:2010qy}.

In order to make contact with large scale structure observations of the matter
distribution, large numerical simulations are usually performed, where a very
specific initial condition is assumed for the primordial spectrum of inhomogeneities
and the evolution is done solving the Newtonian dynamics numerically. In most cases, 
the matter content is just cold dark matter falling into gravitational potential 
wells set in by inflation, although some simulations have included 
also baryons as well as hot dark mater, neutrinos, radiation, and astrophysical feedbacks.

This conceptually simple recipe yields results which
are in good agreement with the matter distribution we observe in 
the sky on large scales, and can be used to constrain our model of the Universe
and determine some of the parameters of the Standard Model of Cosmology.
However, some have argued (see e.~g.~\cite{Rasanen:2010} and references therein) that the late
stages of gravitational collapse, and structure formation in the universe require
a fully relativistic numerical description in order to capture specific 
signatures of the strong non-linear dynamics of general relativity, and make a correct
treatment of features even with sizes comparable to the Hubble radius, see also~\cite{HwangNoh,Mattsson,Bremer}.

In this paper we have tested the validity of the Newtonian approximation for structure formation
in the context of an inhomogeneous model whose fully non-linear dynamics can be 
solved exactly using the Einstein equations~\cite{LTBvoid,GBH:2008a}. We start with
an Einstein-deSitter model at a high redshift, and we test it under various 
initial conditions: i) We first include just a large void of fixed size and a 
small initial amplitude, and we follow the non-linear growth of the void's depth 
and size; ii) We then add Cold Dark Matter (CDM) with a Gaussian random field
distribution based on inflation ($n_s=1$ and $\sigma_8=0.9$), with the weighted
matter-baryon transfer function included to account for the Baryon Acoustic 
Oscillations (with $f_\textrm{gas}=0.14$ and $\Omega_{\rm tot}=1$ consistent with WMAP-7yr),
and follow the growth of both the void and the matter power spectrum.
We have confirmed that our numerical simulations follow the exact 
solution of the LTB background Einstein equations at all scales (radii), except
for very extreme cases, where shell crossing occur in models with large scale
structure fluctuations (shell crossing does not occur though if we take a pure void).
This seem to suggest that the Newtonian  approximation for gravitational collapse
is perfectly valid, even for gigaparsec-sized voids as empty as $\Omega_M=0.02$
at the centre at $z=0$, which corresponds to density contrasts 
of order 1 with respect to the asymptotic EdS model. We obtain very
good matches to both the density and the velocity profile for matter moving in
such LTB backgrounds. We also check that the non-linear evolution gives rise
to well differentiated Hubble rates along the line of sight and transverse
directions, in perfect agreement with the exact relativistic solutions. 
Moreover, we can follow the evolution of the matter density contrast as a function of the
scale factor and find that it evolves as one would expect for an open universe
with the value of $\Omega_M(r)$ corresponding to the local 
position within the void.

\section{Lema\^itre-Tolman-Bondi void models}\label{sec:LTB}

The Lema\^itre-Tolman-Bondi model describes general spherically symmetric space-times and 
can be used as a toy model for describing large voids in the universe. The metric is given by
$$ds^2 = - dt^2 + \frac{A'^2(r,t)\,dr^2}{1-k(r)} + A^2(r,t)\,(d\theta^2 + \sin^2\theta d\phi^2)\,,$$
with a spherically symmetric matter source with negligible pressure,
$T^\mu_{\ \nu} = - \rho_M(r,t)\,\delta^\mu_0\,\delta^0_\nu$. Since we have different radial 
and angular scale factors, we also define a transverse and longitudinal Hubble rates as
$H_T\equiv \dot A / A$, and $H_L\equiv \dot A' / A'$, where dots and primes 
denote $\partial_t$ and $\partial_r$, respectively. Integrating the Einstein
equations for this metric one finds the $r$-dependent transverse Hubble rate
$$\frac{H_T^2(r,t)}{H_0^2(r)} = \Omega_M(r)\left(\frac{A_0(r)}{A(r,t)}\right)^3
+\Omega_K(r)\left(\frac{A_0(r)}{A(r,t)}\right)^2\,,$$
where we have fixed the gauge by setting $A_0(r)=r$ and $\Omega_K(r)=
1-\Omega_M(r)$. For fixed $r$ the above 
equation is equivalent to the Friedmann equation, and has an exact
parametric solution, see Ref.~\cite{GBH:2008a}.

In general, LTB models are uniquely specified by the two functions
$H_0(r)$ and $\Omega_M(r)$, but
to test them against data we have to parameterize the functions, to
reduce the degrees of freedom to a discrete set of parameters.  For
simplicity in this paper we will use the constrained GBH model
\cite{GBH:2008a} to describe the void profile. First of all, it uses a
minimum set of parameters to make a simple void profile, and secondly,
we impose that the time to Big Bang should be constant.  We have made
this choice, because models with an inhomogenous Big
Bang would contain a mixture of growing and decaying modes, and
consequently the void would not disappear at high redshift, making
them incompatible with the Standard Big Bang
scenario~\cite{Zibin:2008a}.  If we only consider constrained LTB
models, then at high redshifts and/or at large distances the central
void is reduced to an insignificant perturbation in an otherwise
homogeneous universe described by an FRW metric, and physical results
for the early universe derived for FRW space-times still hold, even
though we are considering an LTB space-time.
The second condition gives a relation between $H_0(r)$
and $\Omega_M(r)$, and hence constrain the models to one free
function, and a proportionality constant describing the overall
expansion rate. Our chosen model is thus given 
by~\cite{GBH:2008a,GBH:2008b}
\begin{eqnarray}\nonumber
\Omega_M(r) &=&1+ \Big(\Omega_{\rm in} -1\Big) 
\left({1 - \tanh[(r - r_0)/2\Delta r]\over1 + \tanh[r_0/2\Delta r]}\right) \\
\nonumber H_0(r) &=& H_0\left[{1\over \Omega_K(r)} -
{\Omega_M(r)\over\sqrt{\Omega_K^3(r)}}\ {\rm sinh}^{-1}
\sqrt{\Omega_K(r)\over\Omega_M(r)}\right] 
\label{H20}
\end{eqnarray}
where we have assumed that space is asymptotically flat, $\Omega(\infty)=~1$. The model has
then only four free parameters: The overall expansion rate $H_0$, the
underdensity at the centre of the void $\Omega_{\rm in}$, the size of
the void $r_0$, and the transition width of the void profile $\Delta
r$. For more details on the model see Ref.~\cite{GBH:2008a}.

\section{Linear Perturbation Theory}

We still do not have a complete linear perturbation theory for LTB models.
The main difficulty is that since the background is inhomogeneous we cannot
split the perturbations into independent equations for the scalar, vector and tensor modes. 
In LTB models the equations for these modes appear as coupled partial differential 
equations~\cite{Zibin:2008a,Clarkson:2009sc}.
In particular, the scalar modes couple to the tensor shear modes at first order, which act as source for the
scalar mode via the background shear. However, in the case that the latter is small, like in the
models we have been describing in our previous works~\cite{GBH:2008c}, we can ignore this source and solve
{\em exactly} the perturbation equation for the scalar mode $\Phi$, which in the absence of 
anisotropic matter stresses is equal to the curvature mode $\Psi$. In this approximation the equation becomes
\begin{equation}
\ddot\Phi(r,t) + 4H_T(r,t)\dot\Phi(r,t) - \frac{2k(r)}{A^2(r,t)}\Phi(r,t) = 0\,,
\end{equation}
with the exact solution
\begin{equation}
\Phi(r,t) = \Phi_0(r,0)\,{}_2\!F_1\left[1,2,\frac{7}{2},\Big(1-\Omega_M^{-1}(r)\Big)\frac{A(r,t)}{r}\right]\,.
\end{equation}
We note that, strictly speaking, this solution is only exact when ignoring the tensor coupling, and
considering angular transverse modes, but turns out to be a very good approximation.
In that same approximation (negligible background shear), the density contrast of matter is
proportional to the scalar metric perturbation,
\begin{equation}\label{denscontrast}
\delta(r,t) = \delta_0(r) \frac{A(r,t)}{r}\frac{\Phi(r,t)}{\Phi_0(r,0)}\,,
\end{equation}
where $\delta_0(r)$, up to a normalization factor, can be determined under the assumption that the
small scale matter perturbations in the early universe decouple from the void, giving
$\delta_0(r) \propto r / A(r,t_{\textrm early})$.
It is this function which we will try to compare with the simulations described in the next section.

\section{Numerical simulations}

To test the validity of N-Body codes in
describing gigaparsec sized voids, and to follow the evolution and
formation of structure in such models, we have modified the {\tt 2LPT}
initial condition generator~\cite{Crocce:2006ve} to set up an N-Body
simulation of a void for the {\tt Gadget2} code \cite{Gadget1} where
the displacements and velocities of the particles are found using
second order Lagrangian perturbation theory~\cite{Crocce:2006ve}. 
Starting with a standard transfer function for the total matter content 
in a flat Einstein-deSitter model we construct initial conditions
for the gravitational potential in $k$-space $\Phi^i_{\bf k}$. Then
we find the gravitational potential of a void $\Phi^v_{\bf k}$ using the
analytical solution, by interpolating the density out on the particle grid,
and then Fourier transforming it. Now that the total potential
$\Phi_{\bf k} = \Phi^i_{\bf k} + \Phi^v_{\bf k}$ is known, the {\tt 2LPT}
code proceeds unchanged from the original version. Once the initial
conditions have been set up we use the public domain version of the
{\tt Gadget2} code in pure tree-mode to run the simulation (see table
\ref{tab:sims} for an overview of the simulations) \footnote{Running Gadget
in PM-mode, gives an unphysical imprint both at the edge an at the center
of the void, possibly due to the spherical symmetry of the problem.}.

It is not evident that N-body simulations can be used to describe large scale
LTB models, and therefore a significant effort has gone into validating that indeed
we reproduce the expected theoretical behaviour.
To test the code we have used different starting redshifts
($z_{\rm start}$=24, 49, 99, and 199) to check explicitly that the code is started
at high enough redshifts, such that the displacements of the particles
are much smaller than the inter-particle distance, and that the void can be treated
as a linear perturbation, which at first order does not interact with the small
scale fluctuations from the power spectrum. We have used different
resolutions (simulations $\mathcal{S}$24 and $\mathcal{H}$ -- see table \ref{tab:sims})
to test that the cosmological large scale structure is adequately resolved,
we have tested that the void does not interact too much with
mirror images of itself by changing the physical box size from $L$=2400
to $L$=3600 Mpc $h^{-1}$ (simulations $\mathcal{S}$49 and $\mathcal{L}$), and we have
checked that to first order the small scale fluctuations do not back react significantly on
the void, by running with and without matter perturbations (simulations $\mathcal{S}$49
and $\mathcal{V}$). 
Apart from the numerical tests, we have simulated a
representative set of realistic void models varying the transition
length $\Delta r / r_0$ and central underdensity $\Omega_{\rm in}$
(see table~\ref{tab:sims}).
The majority of the simulations use a GBH model with
$\Omega_{\rm in}=0.25$, and $\Delta r/r_0=0.3$, but we have also run other
simulations with $\Omega_{\rm in}=0.125$,
$\Omega_{\rm in}=0.0625$, $\Omega_{\rm in}=0.0208$, and
$\Delta r/r_0=0.1$ and $\Delta r/r_0=0.5$.

\begin{table}
\begin{center}
\begin{tabular}{c|c|c|c|c|c}
\hline \hline
Name & $z_{\rm start}$ & $\Omega_{\rm in}$ & $\Delta r/r_0$ & \#particles & Comments \\
\hline

$\mathcal{H}$ & 24 & 0.25 & 0.3 & $960^3$ & High res sim \\
$\mathcal{V}$ & 49 & 0.25 & 0.3 & $512^3$ & Void alone \\
$\mathcal{S}24$ & 24 & 0.25 & 0.3 & $512^3$ & Void + matter \\
$\mathcal{S}49$ & 49 & 0.25 & 0.3 & $512^3$ & Void + matter \\
$\mathcal{S}99$ & 99 & 0.25 & 0.3 & $512^3$ & Void + matter \\
$\mathcal{S}\Omega$125 & 49 & 0.125 & 0.3 & $512^3$ & Void + matter \\
$\mathcal{S}\Omega$063 & 49 & 0.0625 & 0.3 &  $512^3$& Void + matter \\
$\mathcal{S}\Omega$021 & 199 & 0.0208 & 0.3 & $512^3$ & Void + matter \\
$\mathcal{S}\Delta$01 & 49 & 0.125 & 0.1 & $512^3$ & Void + matter \\
$\mathcal{S}\Delta$05 & 49 & 0.125 & 0.5 & $512^3$ & Void + matter \\
$\mathcal{L}$ & 49 & 0.25 & 0.3 & $768^3$ & $L$=3600 Mpc $h^{-1}$ \\
\hline \hline
\end{tabular} 
\end{center}
\caption{Overview of the simulations. All have been performed
with a void of radius $r_0$=1100 Mpc = 473 Mpc  $h^{-1}$, and
with an asymptotic Hubble parameter of $h_{\infty} = 0.43$.
The standard box size is $L$=2400 Mpc  $h^{-1}$, and the
particle mass is $M_{\rm part}\!\!=2.8\times10^{13} M_\odot h^{-1}$
($M_{\rm part}\!\!=4.3\times10^{12} M_\odot h^{-1}$ for $\mathcal{H}$).
Everywhere we have used a smoothing length of 56 kpc $h^{-1}$
(except for $\mathcal{H}$, where it has been appropriately
rescaled).} \label{tab:sims}
\end{table}

\section{Analysis and results}

The results in this paper show the concordance between the simulations and the theoretical 
predictions. In order to check this we use the highest resolution simulation $\mathcal{H}$
as our reference model and the other simulations to test the limits of the validity of 
N-Body simulations for describing LTB models.
We have subjected our simulation to three different tests, confronting the 
density profile, the Hubble parameter profile ($H_T$ and $H_L$) and the density contrast evolution 
with the corresponding theoretical predictions.

\begin{figure}
\begin{center}
\includegraphics[width=0.48 \textwidth]{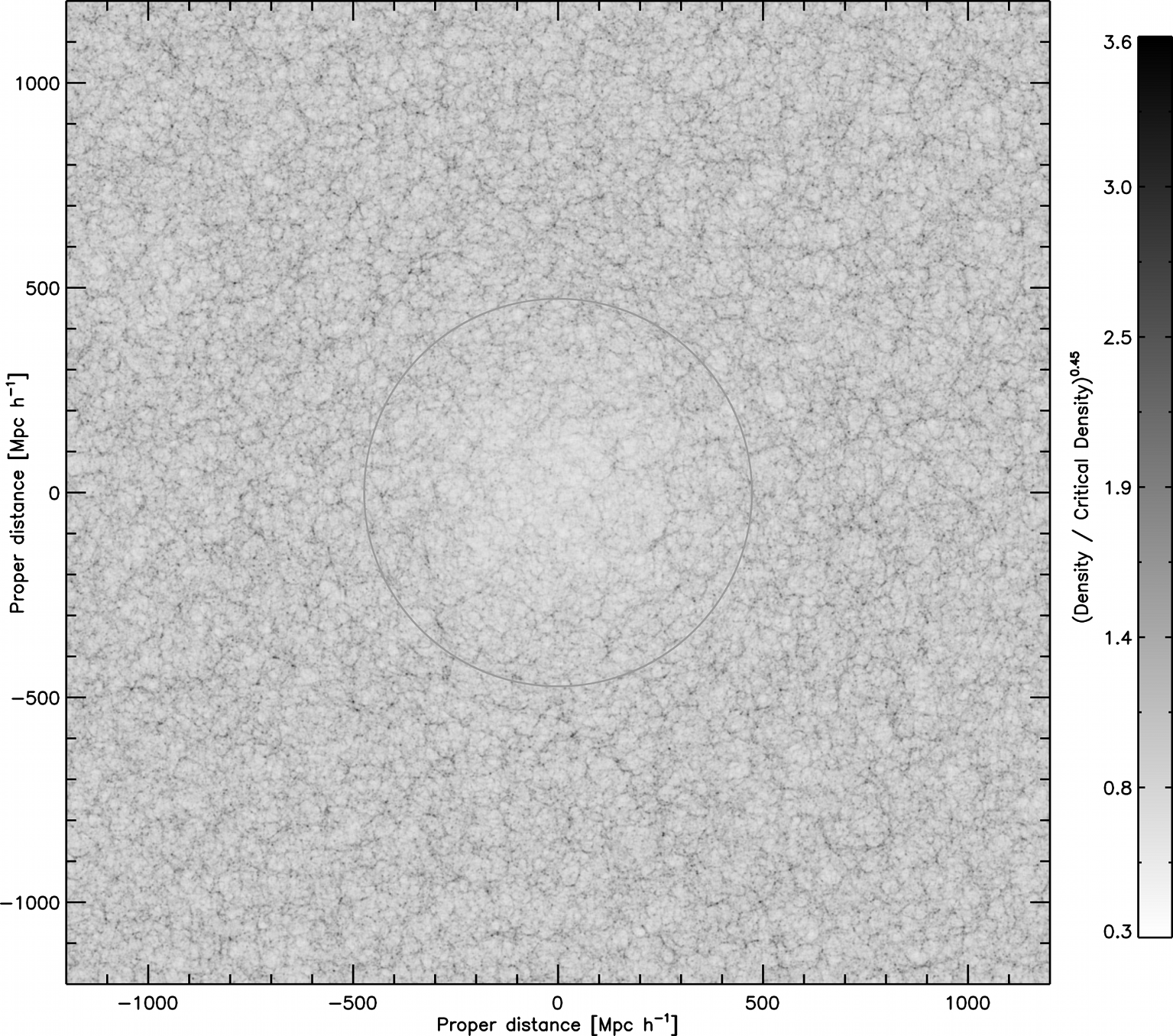}
\caption{The projected matter distribution  at $z=0$ averaged over a 175 Mpc
slice centered on the void of the $960^3$ simulation.
Notice how near the centre of the void, not only the density is lower,
but also there is significantly less structure than outside the void.
The characteristic void size $r_0=473$ Mpc $h^{-1}$ is indicated by the thin circle.}
\label{fig:rho960}
\end{center}
\end{figure}

\subsection{Distances and redshifts in LTB models.} 

{\tt Gadget2} is designed to perform simulations of  FRW universes, and one needs to associate the
comoving radial coordinates in both models. Since we have analyzed the data from {\tt Gadget2} snapshots,
that is, positions and velocities of particles are ``measured'' at constant cosmic time, and all our 
observables are quantities calculated in thin spherical shells, this identification must be done through the
proper radial distance, calculated in both cases as
\begin{equation} \label{eq:dp}
 d_p(r,t) = a(t)\,r_{\rm FRW} = \int_0^{r_{\rm LTB}}\frac{A'(r,t)}{\sqrt{1 - k(r)}}dr\,.
\end{equation}
When the curvature factor $(1 - k(r))^{-1/2}$ is roughly 1, which is the case in the models
under study, one can approximate
\begin{equation}
 d_p(r,t) = a(t)\,r_{\rm FRW} \simeq A(r_{\rm LTB},t) \label{eq:approx}
\end{equation}
for most redshifts. Similarly, when interpreting the results, it is important to 
remember that while the proper cosmological time in the two metrics can readily be identified, the
redshift at equal times are different, i.e.~for $t_{\rm FRW}=t_{\rm LTB}$ the $z_{\rm FRW}$ and
$z_{\rm LTB}$ are different. It is important to emphasize that since we are considering a constrained-GBH
LTB model, the time to Big Bang is homogenous and thus all times at each radial distance are the same,
so each particle in the simulation has a time given by the code: $t_{\rm FRW}=t_{\rm LTB}$.

\subsection{Density profile}

\begin{figure}
\begin{center}
\includegraphics[width=0.45 \textwidth]{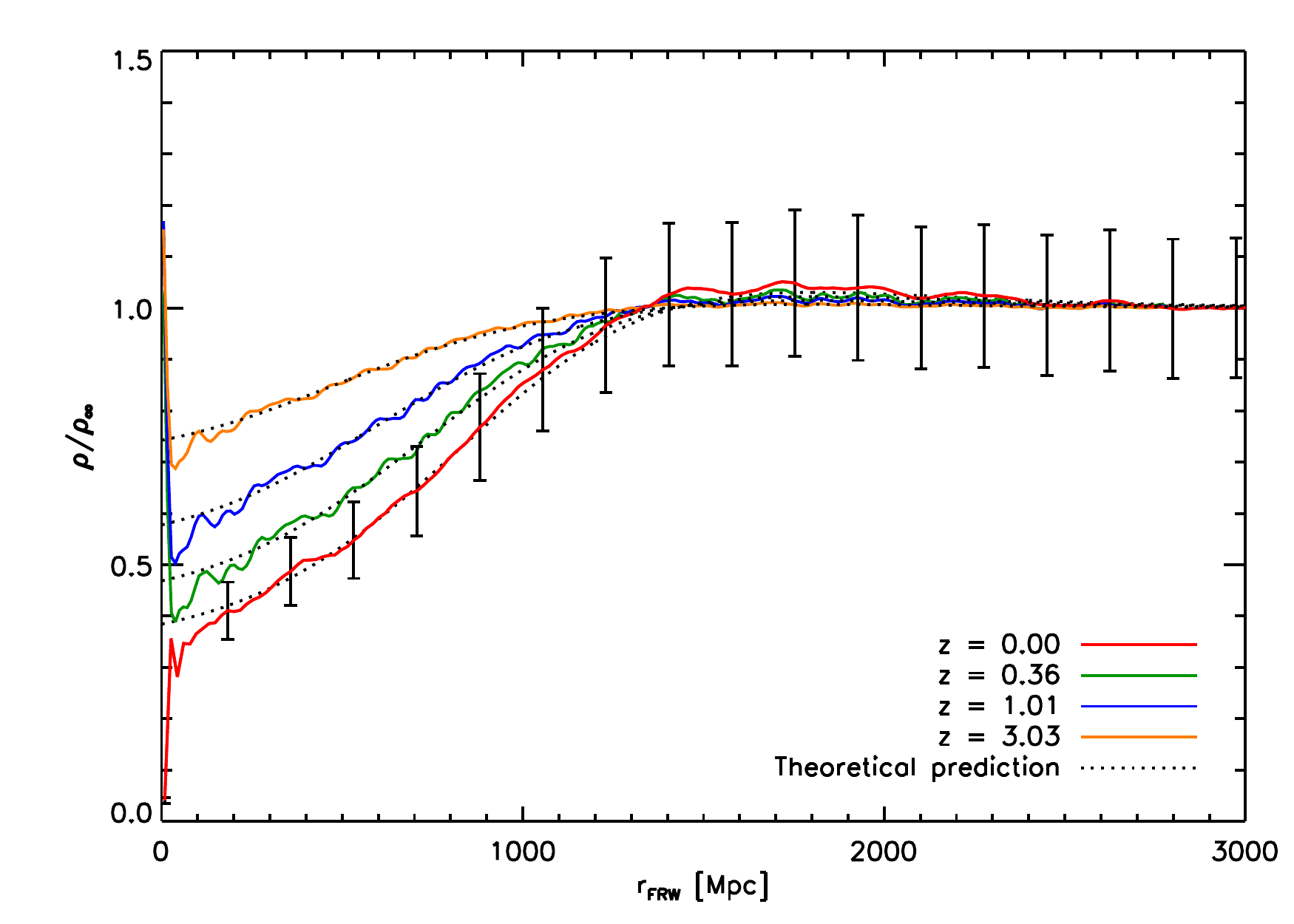}
\caption{Comparison of the density profile of the $\mathcal{H}$ simulation at different redshifts with the 
theoretical curves, as a function of comoving distance $r_{\rm FRW}=(1+z)d_p$ in Mpc.}
\label{fig:avd960snaps}
\end{center}
\end{figure}

We first compare the theoretical density profile of a GBH universe having the desired parameters 
with the corresponding profile obtained from the simulation. The density field is calculated by interpolating
each particle in the box to a grid using a 2$^\textrm{nd}$ order triangular-shaped-cloud
(TSC) technique~\cite{Hockney} (see fig.~\ref{fig:rho960}); then the simulation box is divided into
different spherical bins, and we calculate the average density in each of them thus obtaining the density as
a function of the proper distance $d_p$ (see eq.~\ref{eq:dp}). 
Due to the presence of non-linear inhomogeneities, the error in the
determination of the density profile cannot be directly obtained as the r.m.s.~in each bin, and the error bars
displayed in the figures have been calculated as the r.m.s.~in the analogous $\mathcal{V}$ simulation without
CDM perturbations. The reference simulation $\mathcal{H}$ shows an excellent agreement between theory and
simulation (see fig.~\ref{fig:avd960snaps}), except near the centre of the void, where the particle distribution
is undersampled and shot noise dominated.

\begin{figure}[t]
\begin{center}
\includegraphics[width=0.45 \textwidth]{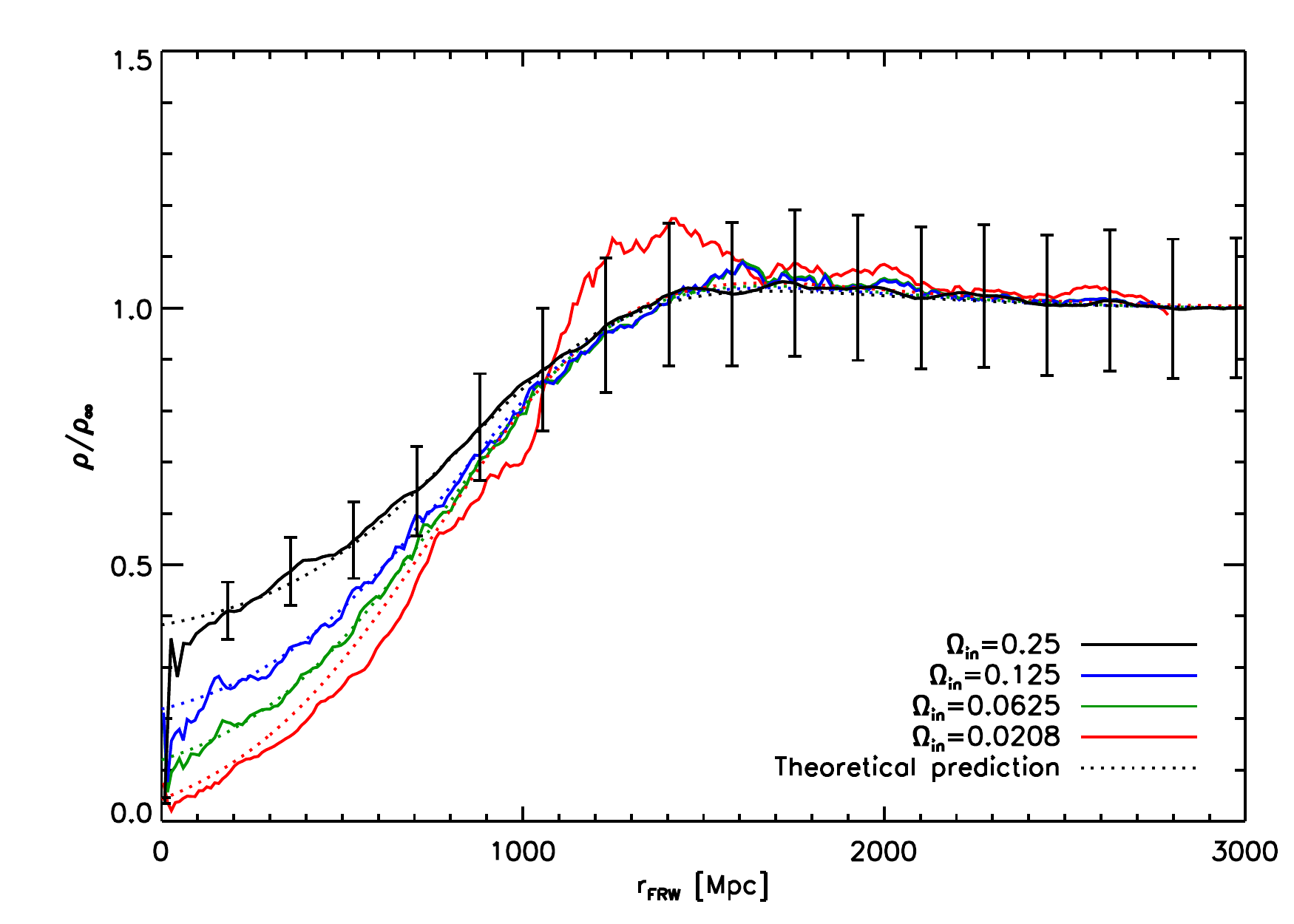}
\includegraphics[width=0.45 \textwidth]{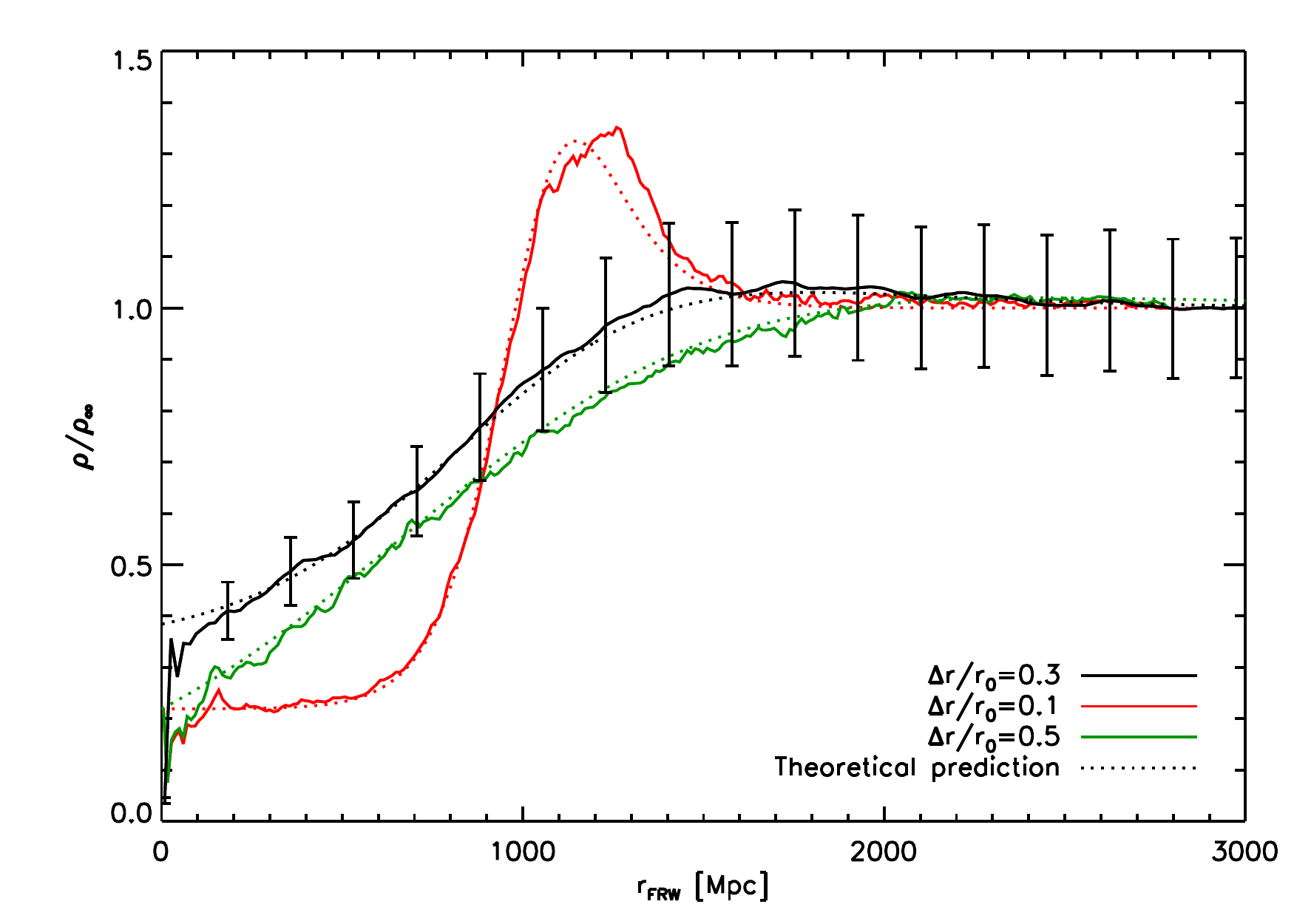}
\caption{Density profiles for different values of $\Omega_{\rm in}$ (top panel) and $\Delta r/r_0$ (lower panel) in comparison with the corresponding theoretical profiles. All curves are plotted at redshift $z=0$.}
\label{fig:avd49_rhos}
\end{center}
\end{figure}

In fig.~\ref{fig:avd49_rhos} we show the density profile for an extended set of models. For most models
the simulations are in excellent concordance with the theory, though for two extremal cases, namely the
emptiest void $\mathcal{S}\Omega$021, and the void with the steepest transition $\mathcal{S}\Omega\Delta$01
we find significant deviations. For $\mathcal{S}\Omega\Delta$01 the discrepancy is not severe, and only present
in the density profile. We speculate that this could be due to under resolution of the transition length or possibly
due to the small-scale perturbations interacting with the large scale void, given that the transition length is only
$\Delta r=47.3$ Mpc $h^{-1}$.

\begin{figure}
\begin{center}
\includegraphics[width=0.45 \textwidth]{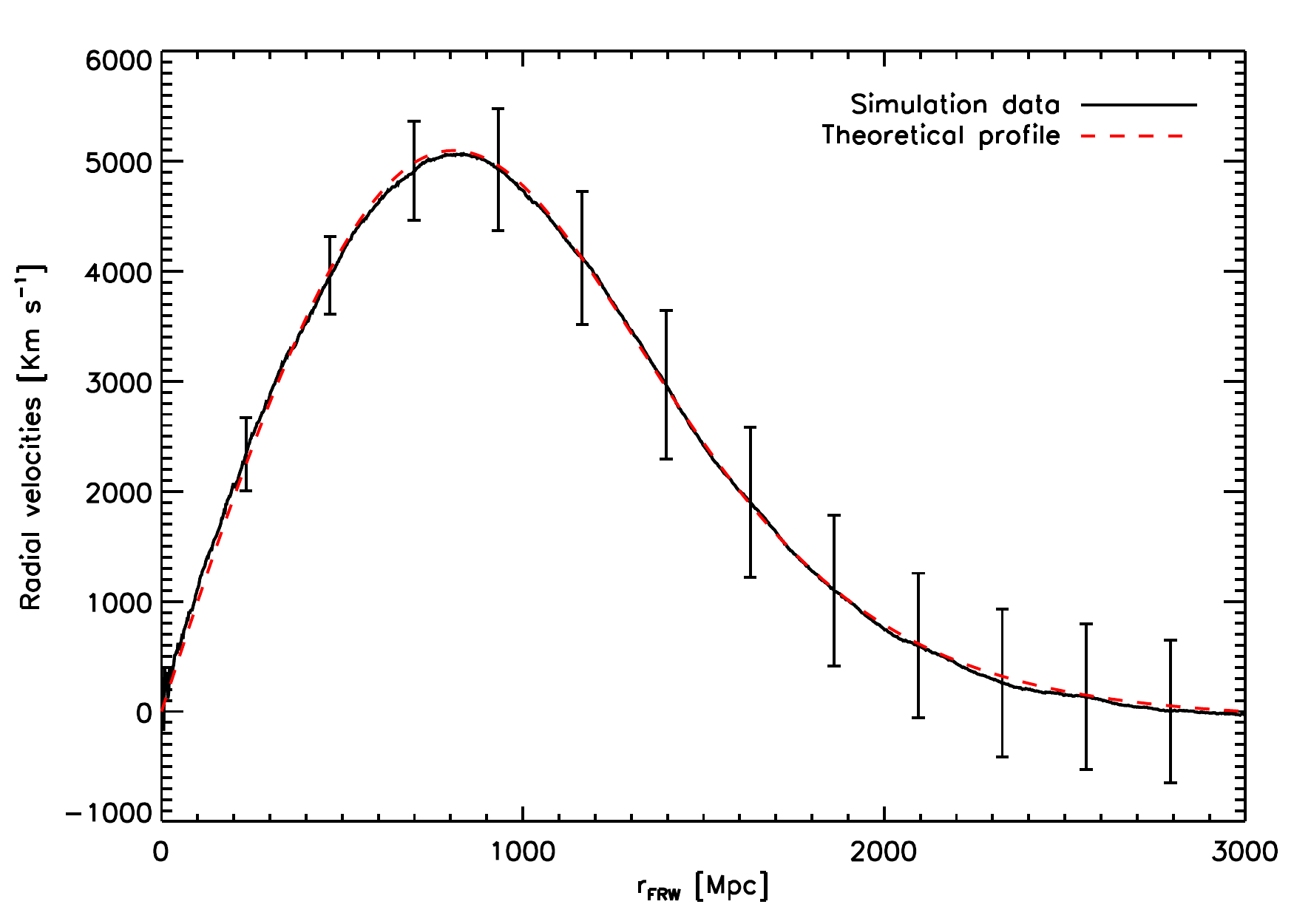}
\includegraphics[width=0.45 \textwidth]{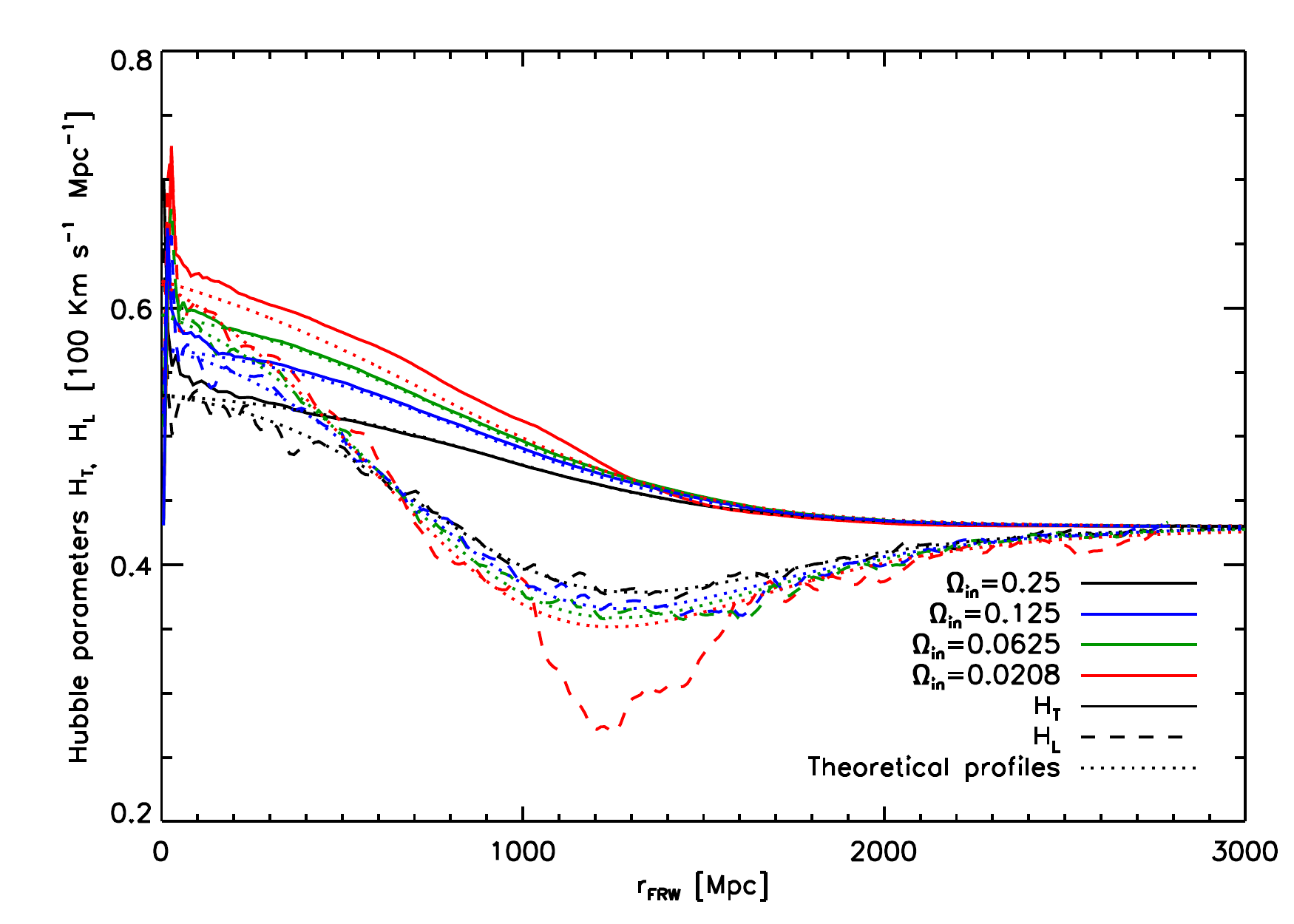}
\caption{The velocity profile for the simulation $\mathcal{H}$ (top panel) and the $H_T$ and $H_L$ 
profiles for different values of $\Omega_{\rm in}$ (lower panel). In both cases, the theoretical profiles
are shown with dotted lines. All curves are plotted at redshift $z=0$.}
\label{fig:H960}
\end{center}
\end{figure}

\subsection{Rates of expansion}

The radial velocity profile can be used to compare against the theoretical predictions for $H_T$ and $H_L$.
The rate of change in the proper distance $\dot{d}_p/d_p$ computed in the rest-frame of the matter should
match each other in the FRW and LTB metric, if the simulations are a valid description of the LTB model.
In the LTB metric matter is at rest, and keeps the same comoving coordinate, while in the FRW metric there
are systematic radial motion, and We have that
\begin{equation}
\frac{d}{dt}d_p^{\textrm{FRW}} = \frac{d}{dt}\left[ a\,r_{matter} \right] = d_p \left[ \langle v_r \rangle / r + H_{\infty} \right]\,,
\end{equation}
which can be directly compared to the theoretical LTB result calculated taking the derivative of the
r.h.s.~of eq.~(\ref{eq:dp}). $\langle v_r \rangle$ is calculated as the average radial velocity $v_r$ of
the particles sampled in spherical bins. In the upper panel of fig.~\ref{fig:H960} we see how the theoretical
radial velocities (calculated from $a^{-1}[\dot{d}_p - H_{\infty} d_p]$ ) match the data from $\mathcal{H}$. 
We have found that $\dot{d}_p/d_p$ aproximate $H_T$ very well (see eq.~\ref{eq:approx}),
possibly because the denominator in eq.~(\ref{eq:dp}), $(1-k(r))^{-1/2}$, being time independent, cancel
in the ratio $\dot{d}_p / d_p$. 
Using this approximation in the lower panel of fig.~\ref{fig:H960} we compare a range of models
to theory. Again, the difference with the theoretical graph found near $d_p=0$ is understandable,
we are shot noise dominated, and furthermore the matter perturbations displace the centre of the void slightly,
while at the same time we have a formal singularity at $r=0$ when calculating $\langle v_r \rangle/r $.
From $H_T$ we can extract $H_L$ straightforwardly as :
\begin{equation}
 H_L= \frac{\dot{A}'}{A'} = H_T + \frac{A}{A'}H'_T,
\end{equation}
which  is just $H_L = H_T + r\,H'_T$ at $z = 0$, using $A_0(r) = r$. We stress though, that at $z=0$
this is a derived parameter, and not independent of $H_T(d_p)$. We find that all but the emptiest model
$\mathcal{S}\Omega$021 match well the theoretical predictions. For $\mathcal{S}\Omega$021 the
velocity is consistently higher (and the density lower) inside the void compared to theoretical
predictions, and a density spike is building up near the edge of the void. This could be due to the
very lower density, but it may also be a consequence of the very high starting redshift ($z_{\rm start}=199$),
that was necessary to keep the perturbations linear and the particle displacements acceptable in the
initial condition.

\subsection{Density contrast}

\begin{figure}[!t]
\begin{center}
\includegraphics[width=0.45 \textwidth]{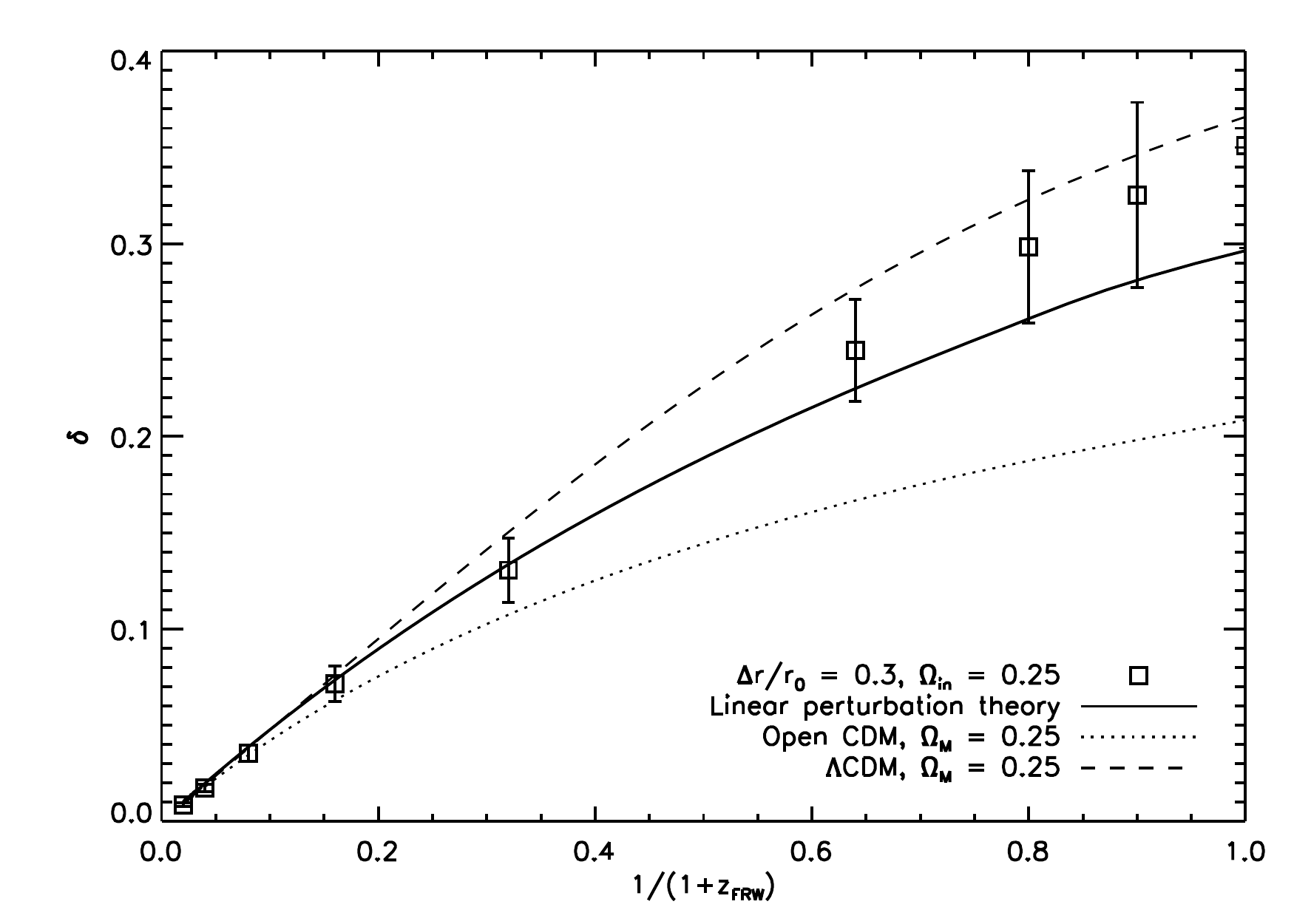}
\caption{Density contrast evolution inside the void at comoving distance $r_{\rm FRW}=(1+z)\,d_p=280$ Mpc, 
for $\mathcal{S}$49, in comparison with the theoretical prediction from perturbation theory 
(full line). We also compare the LTB growth of density perturbations with that of Open CDM 
(dotted line) and $\Lambda$CDM (dashed line). The theoretical curves were normalized to have the
same slope asymptotically in the past, as $a_{\rm FRW} \to 0$. Note that, even though the horizontal axis 
reads $1/(1+z_{\rm FRW})$, this $z_{\rm FRW}$ only determines the cosmic time $t$, since the
density contrast was calculated at a fixed comoving distance, and not in the lightcone.}
\label{figs:deltamed49}
\end{center}
\end{figure}

Another interesting observable to study is the evolution of the density contrast as a function of redshift,
$\delta(z)=\langle (\rho(z)-\bar \rho) /\bar \rho \rangle $. Being (random) fluctuations, we calculate it by finding the
r.m.s.~of $\delta(z)$ in spherical bins as a function of proper distance. The errors in the determination of $\delta$ were
calculated as the standard deviation of the values of $\delta$ calculated in the 8 octants of each spherical bin. 
The results for the simulation $\mathcal{S}$49 can be seen in fig.~\ref{figs:deltamed49}, where we compare the 
density contrast, calculated at a fixed comoving distance $r_{\rm FRW} = (1+z)\,d_p$, as a function of time (expressed 
in terms of redshift), with the predicted one within the simplified linear perturbation theory in LTB described by eq.~(\ref{denscontrast}). We also include, for comparison, the density contrast growth for an open universe, with $\Omega_M=0.25$, and a $\Lambda$CDM, with $\Omega_M=0.25$ and $\Omega_\Lambda=0.75$. It is interesting
to note that the data agree well, within error bars, with both the theoretical prediction in the LTB model and in the 
concordance $\Lambda$CDM, while they differ significantly from an open universe with the same matter density.

\begin{figure}[!t]
\begin{center}
\includegraphics[width=0.45 \textwidth]{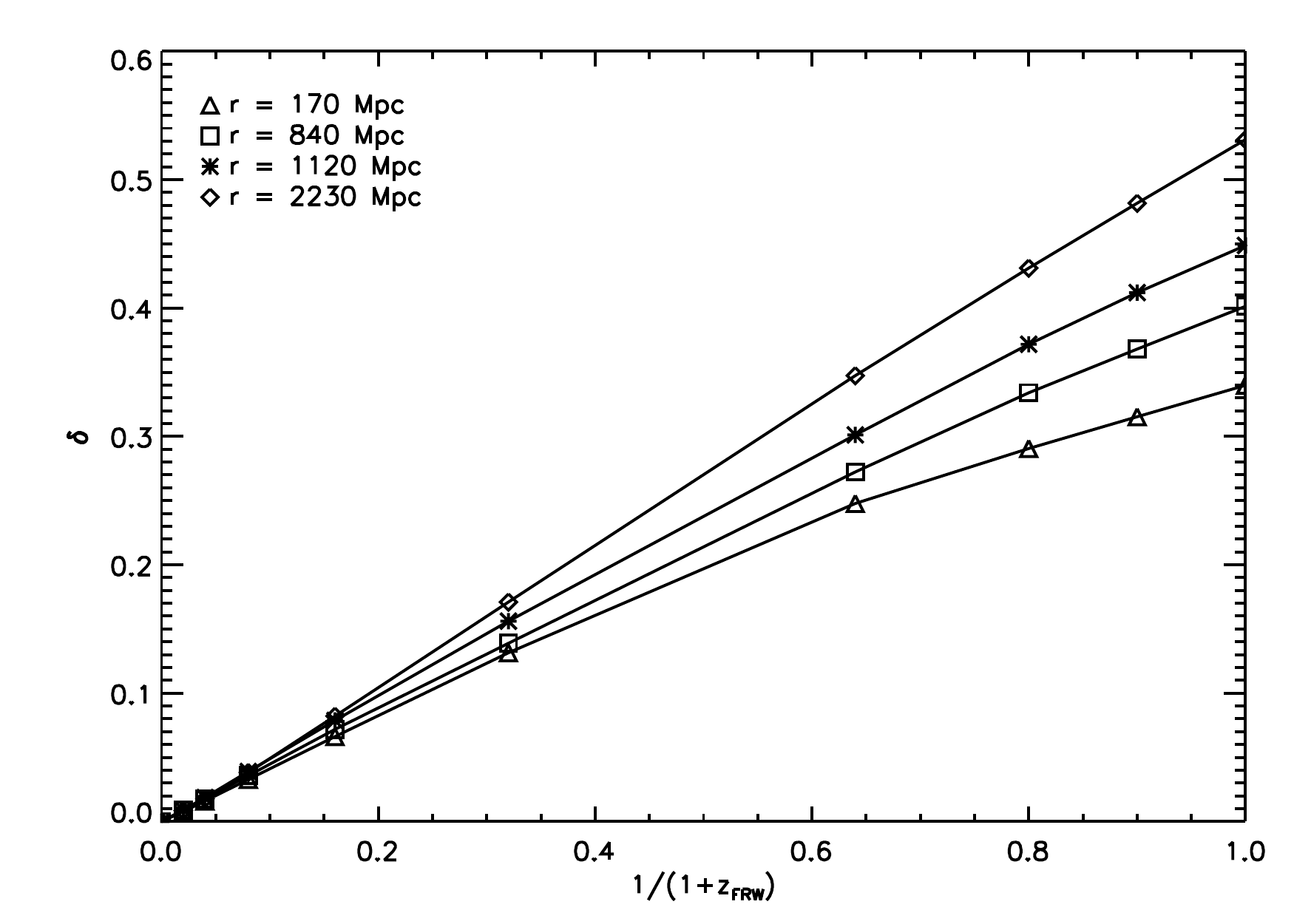}
\caption{Density contrast evolution at different fixed comoving distances for $\mathcal{S}$49 simulation, as a
function of the FRW scale factor, like in Fig.~5. It is easy to distinguish between the contrast growth in a 
background with $\Omega\sim 1$ at large distances and with a lower $\Omega$ near the void center.\\[6.2ex]}
\label{figs:delta49multi}
\end{center}
\end{figure}

In order to better understand these differences, we have also studied the evolution of the density contrast at several 
distances from the center of the void. The results are shown in fig.~\ref{figs:delta49multi}. Two clearly different zones 
can be distinguished: while the growth is proportional to $(1+z)^{-1}$ for large comoving distance, i.e. $\Omega\sim 1$ 
outside the void and $a_{\rm FRW}=(1+z)^{-1}$, the growth is significantly slower (as would occur in an open FRW 
universe), for small distances.

\section{Conclusions}

We have studied for the first time the non-linear evolution of structure formation in large-void LTB 
models within an asymptotic Einstein-deSitter (EdS) universe. By initiating large N-body simulations at high 
redshifts, we have been able to follow the non-linear gravitational collapse of matter structures 
in the presence of an underdense void that starts with a density contrast of order $\delta_m\sim10^{-3}$ at photon decoupling 
(where the matter perturbations have $\delta_m\sim10^{-5}$). We find that using a standard N-body code, the nonlinear
growth of the  void underdensity follows the exact analytical solution of the Einstein equations, even for very deep voids with
$\Omega_{M}=0.06$ at the center, and thus with density contrasts of order one with respect to the asymptotic 
EdS universe. Moreover, the transverse and longitudinal rates of
expansion agree with the theoretical expectations, giving us confidence that the simulations are tracing the
full non-linear gravitational collapse in this non-perturbative LTB background. This is furthermore evidence that N-body
codes give a credible and precise description of the standard $\Lambda$CDM model, where the voids are of much
lesser size, and no general relativistic corrections are needed to describe the large scale evolution.

We have also studied the evolution of the matter density contrast in such a non-trivial background, and found an
analytical solution to the approximate equations for the growth of perturbations  in the limit of negligible background
shear, and shown that the numerical and analytical results are in good agreement. Moreover, the comparison with OCDM
and $\Lambda$CDM shows that the density contrast growth for our LTB models is very close within errors to that of the
concordance $\Lambda$CDM models suggested by WMAP-7yr~\cite{Komatsu:2010fb}.

From our non-linear LTB N-body simulations we can extract predictions for observations of large scale structure,
via the two-point angular correlation function, the angular power spectra, the growth of structure, and the density contrast.
Our models therefore give the possibility of using both current and future observations of the large scale structure 
(such as DES~\cite{DES}, EUCLID~\cite{Euclid} and PAU~\cite{PAU}), to constrain LTB models that already provides 
a viable fit to current observations of the geometry of the Universe.

\section*{Acknowledgements} 

We thank the anonymous referee for very helpful comments on the manuscript. 
TH acknowledges support from the Danish Natural Science Research Council.
JGB thanks the Institute de Physique Th\'eorique de l'Universit\'e de Gen\`eve for their generous hospitality
during his sabbatical in Geneva. DAM acknowledges support from a JAE-Predoc contract. We also thank
the Benasque Center for Science Pedro Pascual, where this work was partially developed.
This work is supported by the Spanish MICINN under Project No. AYA2009-13936-C06-06, 
the CAM project ``HEPHACOS'' Ref. S2009/ESP-1473, and by the EU FP6 Marie Curie Research and 
Training Network ``Universe Net'' Ref. MRTN-CT-2006-035863. 
Computer time was provided by the Danish Center for Scientific Computing.


\begin{thebibliography}{99}

\bibitem{Komatsu:2010fb}
  E.~Komatsu {\it et al.},
  arXiv:1001.4538 [astro-ph.CO].

\bibitem{Gott:2003pf}
  J.~R.~I.~Gott {\it et al.},
  {\em Astrophys.\ J.\ }  {\bf 624}, 463 (2005),
  arXiv:astro-ph/0310571.

\bibitem{LTBvoid}
N. Mustapha, C. Hellaby \& G.~F.~R. Ellis,
{\em Mon. Not. Roy. Astron. Soc.} {\bf 292}, 817 (1997);
M.-N. Celerier, {\em Astron. Astrophys.} {\bf 353}, 63 (2000);
K. Tomita, {\em Mon. Not. Roy. Astron. Soc.} {\bf 326}, 287 (2001);
J.~W. Moffat, {\em JCAP} {\bf 0510}, 012 (2005);
H. Alnes, M. Amarzguioui \& O. Gron, {\em Phys. Rev. D} {\bf 73}, 083519 (2006);
D. Garfinkle, {\em Class. Quant. Grav.} {\bf 23}, 4811 (2006);
K. Enqvist \& T. Mattsson, {\em JCAP} {\bf 0702}, 019 (2007);
K. Enqvist, {\em Gen. Rel. Grav.} {\bf 40}, 451 (2008);
T. Mattsson, arXiv:0711.4264 [astro-ph];
D.~L. Wiltshire, arXiv:0712.3984 [astro-ph].

\bibitem{GBH:2008a}
  J.~Garcia-Bellido and T.~Haugboelle,
  {\em JCAP} {\bf 0804}, 003 (2008),
  arXiv:0802.1523 [astro-ph].

\bibitem{Cruz}
M.~Cruz {\it et al.},
{\em Mon. Not. Roy. Astron. Soc.} {\bf 369} 57 (2006);
 {\em Astrophys. J.} {\bf 655} 11 (2007);
{\em Mon. Not. Roy. Astron. Soc.} {\bf 390} 913 (2008).

\bibitem{Frith:2003}
W.~J. Frith, G.~S. Busswell, R.~Fong, N.~Metcalfe \& T.~Shanks, 
{\em Mon. Not. Roy. Astron. Soc.} {\bf 345} 1049 (2003)
  
\bibitem{Granett:2008}
B.~R. Granett, M.~C. Neyrinck \& I. Szapudi,
  {\em Astrophys. J. Lett.} {\bf 683}, L99 (2008).

\bibitem{Tully:2007tp}
  R.~B.~Tully,
  arXiv:0708.0864 [astro-ph].

\bibitem{Kashlinsky:2008a}
  A.~Kashlinsky, F.~Atrio-Barandela, D.~Kocevski and H.~Ebeling,
  {\em Astrophys. J. Lett.} {\bf 686}, L49 (2008).

\bibitem{GBH:2008b}
  J.~Garcia-Bellido and T.~Haugboelle,
  JCAP {\bf 0809}, 016 (2008),
  arXiv:0807.1326 [astro-ph];

\bibitem{GBH:2008c}
  J.~Garcia-Bellido and T.~Haugboelle,
  JCAP {\bf 0909}, 028 (2009),
  arXiv:0810.4939 [astro-ph].

\bibitem{Biswas:2010xm}
  T.~Biswas, A.~Notari and W.~Valkenburg,
  arXiv:1007.3065 [astro-ph.CO].

\bibitem{Clarkson:2010ej}
  C.~Clarkson and M.~Regis,
  arXiv:1007.3443 [astro-ph.CO].

\bibitem{Moss:2010jx}
  A.~Moss, J.~P.~Zibin and D.~Scott,
  arXiv:1007.3725 [astro-ph.CO].

\bibitem{Yoo:2010qy}
  C.~M.~Yoo, K.~i.~Nakao and M.~Sasaki,
  JCAP {\bf 1007}, 012 (2010),
  arXiv:1005.0048 [astro-ph.CO];
  arXiv:1008.0469 [astro-ph.CO].


\bibitem{Rasanen:2010}
R{\"a}s{\"a}nen, S., {\em Phys.~Rev.~D} {\bf 81}, 103512 (2010).

\bibitem{HwangNoh}
  J.~Hwang and H.~Noh,
  {\em Gen.\ Rel.\ Grav.}\  {\bf 38}, 703 (2006); 
  {\em Mon. Not. Roy. Astron. Soc.} {\bf 367}, 1515 (2006).

\bibitem{Bremer}
  M.~N.~Bremer, J.~Silk, L.~J.~M.~Davies and M.~D.~Lehnert,
  arXiv:1004.1178 [astro-ph.CO].

\bibitem{Mattsson}
  M.~Mattsson and T.~Mattsson,
  JCAP {\bf 1010}, 021 (2010),
  arXiv:1007.2939 [astro-ph.CO].

\bibitem{Zibin:2008a}
J.~P. Zibin, {\em Phys. Rev. D} {\bf 78}, 043504 (2008).

\bibitem{Clarkson:2009sc}
  C.~Clarkson, T.~Clifton and S.~February,
  JCAP {\bf 0906}, 025 (2009),
  arXiv:0903.5040 [astro-ph.CO].

\bibitem{Crocce:2006ve}
M.~Crocce, S.~Pueblas \& R.~Scoccimarro,
{\em Mon. Not. Roy. Astron. Soc.} {\bf 373} 369 (2006).

\bibitem{Gadget1}
V.~Springel, N.~Yoshida \& Simon D.~M. White, {\em New Astron.} {\bf 6}, 79, (2001);
V.~Springel, {\em Mon. Not. Roy. Astron. Soc.} {\bf 364}, 1105 (2005).

\bibitem{Hockney}
R.W. Hockney \& J.W. Eastwood, ``Computer Simulations Using Particles",
Eds. Taylor \& Francis (1989).

\bibitem{DES} \url{http://www.darkenergysurvey.org}.

\bibitem{Euclid} \url{http://sci.esa.int/euclid/}

\bibitem{PAU}
N.~Benitez {\it et al.}, {\em Astrophys. J.} {\bf 691}, 241 (2008),
  arXiv:0807.0535 [astro-ph.CO];
R.~Casas {\it et al.}, ``The PAU Camera,''
  Proc.\ SPIE Int.\ Soc.\ Opt.\ Eng.\  {\bf 7735}, 773536 (2010).



\end{thebibliography}
\end{document}